\documentclass[prl,floatfix,twocolumn,superscriptaddress,showpacs]{revtex4}
\usepackage{graphicx}
\begin{document}

\title{Distribution of the $\mathbf S$-matrix in chaotic microwave cavities with direct processes
and absorption}

\author{U.~Kuhl}
\affiliation{Fachbereich Physik, Philipps-Universit\"at Marburg, Renthof 5,
D-35032 Marburg, Germany}

\author{M.~Mart\'{\i}nez-Mares}
\affiliation{Departamento de F\'{\i}sica, UAM-Iztapalapa,
Av. San Rafael Atlixco 186, Col. Vicentina,
09340 M\'exico D. F., M\'exico}

\author{R.A.~M\'endez-S\'anchez}
\affiliation{Centro de Ciencias F\'{\i}sicas, Universidad Nacional Aut\'onoma de
M\'exico, A.P 48-3, 62210, Cuernavaca, Mor., M\'exico}

\author{H.-J.~St\"ockmann}
\affiliation{Fachbereich Physik, Philipps-Universit\"at Marburg, Renthof 5,
D-35032 Marburg, Germany}

\date{\today}

\begin{abstract}
We quantify the presence of direct processes in the $S$-matrix of chaotic microwave
cavities with absorption in the one-channel case. To this end the full distribution
$P_S(S)$ of the $S$-matrix, i.e. $S=\sqrt{R}e^{i\theta}$, is studied in cavities with
time-reversal symmetry for different antenna coupling strengths $T_a$ or direct
processes. The experimental results are compared with random-matrix calculations
and with numerical simulations based on the Heidelberg approach including absorption.
The theoretical result is a generalization of the Poisson kernel.
The experimental and the numerical distributions are
in excellent agreement with random-matrix predictions for all cases.
\end{abstract}

\maketitle

Random-matrix theory has been successfully applied to many different scattering
systems in several branches of physics ranging from quantum mechanics and mesoscopic
physics to sound and microwave systems \cite{Bee97,Stoe99,Mel04}. The
only precondition is the existence of chaotic ray dynamics of the system.
Nevertheless, the connection between the statistical properties of scattering
and the underlying chaos is not straightforward since scattering may involve two
time scales, namely a prompt and a delayed response. Prompt or direct processes are
those in which the waves pass through the interaction region without a significant
delay. In the equilibrated or delayed processes the waves suffer several reflections
inside the interaction region. The delayed processes are
usually studied with techniques of random matrix theory,
whereas the direct processes are described in terms of the average of the $S$-matrix.

Starting from the pioneering work of L\'opez, Mello and Seligman
\cite{Lop81}, there are several theoretical works addressing the
statistical distributions of the $S$-matrix with imperfect coupling or direct
processes
\cite{Sel83,Mel85,Fri85,Dor92,Bro95,Sav01,Acg03}.
This distribution is known in literature as the Poisson kernel. In
the one-channel case with no absorption it is possible to parameterize the $S$-matrix
as $S(E)=e^{i\theta(E)}$, and the Poisson kernel reads
\begin{equation}
p(\theta)=\frac{1}{2\pi}
\frac{1-\left|\langle S\rangle\right|^2}{\left| S-\langle S \rangle \right|^2},
\label{Eq:PoissonKernel}
\end{equation}
where $\langle S \rangle$ is the ensemble (or energy) average of $S(E)$. For
$\langle S\rangle=0$ the distribution of the phase $\theta(E)$ is uniform between 0 and
$2\pi$, i.e., $S(E)$ is uniformly distributed on the unitary circle, in agreement with
the circular ensembles of random-matrix theory. Eq.~(\ref{Eq:PoissonKernel}) means
that the $S$-matrix distribution of a system including direct processes is fixed
by the average $\langle S\rangle$ exclusively.

Following Brouwer \cite{Bro95}, the Poisson kernel can be interpreted as follows:
For a chaotic system with an attached waveguide with ideal coupling, the distribution
of the $S$-matrix is uniform. If the coupling in the waveguide becomes non-ideal
then the new $S$-matrix is distributed according to the Poisson kernel.
Thus, the Jacobian of the transformation between $S$-matrices of the system with
ideal coupling to the system with non-ideal coupling yields the Poisson
kernel, Eq.~(\ref{Eq:PoissonKernel}).

Unfortunately the microwave experiments cannot be directly compared with Poisson's kernel
Eq.~(\ref{Eq:PoissonKernel}) due to losses or absorption.
When absorption is present, $S$ is a subunitary
matrix. For the one-channel case the $S$-matrix can be
parameterized as
\begin{equation}
S = \sqrt{R} e^{i\theta},
\label{eq:para}
\end{equation}
where $R$ is the reflection coefficient. The coupling between the scattering
channels and the interior region can be
quantified by the transmission coefficient $T_a$ of a barrier
describing the direct processes:
\begin{equation}
T_a = 1 - |\langle S \rangle|^2.
\label{eq:coupling}
\end{equation}
The subindex $a$ will be used to denote the antenna coupling.
For perfect antenna coupling ($T_a=1$) there are some exact results. The phase $\theta$ is
uniformly distributed between 0 and $2\pi$ as before. The distribution $P_{R,0}(R)$
of $R$ is known in the cases of strong ($\gamma\gg 1$) \cite{Mel00} and weak
($\gamma\ll 1$) absorption \cite{Bee01} for systems with and without time
reversal symmetry ($\beta=1,2$).
Throughout this letter the subindex ``0'' refers to perfect coupling.
In case of systems without time
reversal symmetry and a single perfectly coupled channel the distribution $P_{R,0}(R)$ was
calculated for any absorption strength $\gamma$ by Beenakker and Brouwer
\cite{Bee01}, as well as for two perfectly coupled channels in presence of
time-reversal symmetry. For systems with imperfect coupling and absorption the
distributions of the proper time delays and reflection eigenvalues are only known for
$\beta=2$ \cite{Sav03}. More recently, a general distribution of the
reflection eigenvalues has been obtained for a large number of propagating channels,
independent of time-reversal symmetry \cite{Sav04}. Other quantities, such
as probabilities of no return, distributions of Wigner time delay have
also been obtained for systems with absorption \cite{Fyo03,Fyo04}.

From the experimental point of view
the effect of the absorption has been studied on the $S$-matrix
correlation function \cite{Dor90,Lew92}, the
reflection coefficient
\cite{Dor90,Men03}, the
cross-correlation functions of the $S$-matrix~\cite{Sch03},
the transmission coefficient~\cite{Sch01}
and very recently on the impedance matrix~\cite{Hem}.

In this paper measurements are presented which provide clear evidence of direct
processes in chaotic scattering with absorption. In particular the distribution
$P_\theta(\theta)$ of the phase of the subunitary $S$-matrix given in Eq.~(\ref{eq:para})
will be obtained. This corresponds to a generalization of Poisson's kernel,
Eq.~(\ref{Eq:PoissonKernel}), including absorption.
Also the full experimental distribution
$P_S(S)$ of the $S$-matrix in presence of both absorption and direct processes for the
one-channel case is given. Both distribution are related through
$P_\theta(\theta) = \int_{0}^{1}P_S(S)dR$.
We compare our measurements with
numerical simulations based on the Heidelberg approach and with our random-matrix
analytical calculations.

\begin{figure}
\includegraphics[width=\columnwidth]{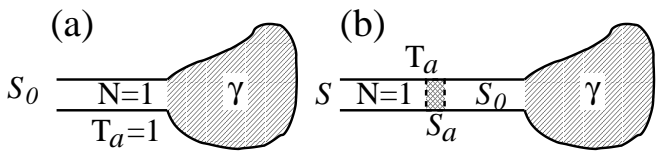}
\caption{Sketch of the model for scattering with direct processes and absorption. In
(a) $S_0$ describes the scattering of a billiard with absorption $\gamma$ but perfect
coupling to the lead ($T_a=1$). In (b) we associate the $S$-matrix $S_a$ given by
Eq.~(\ref{Sbarrier}) to the barrier that describes the non-ideal coupling. The resulting
$S$-matrix of the system with absorption $\gamma$ and coupling $T_a$
can be written in terms of $S_0$ and $S_a$.}
\label{fig:sketch}
\end{figure}

\begin{figure}
\includegraphics[height=1.9cm]{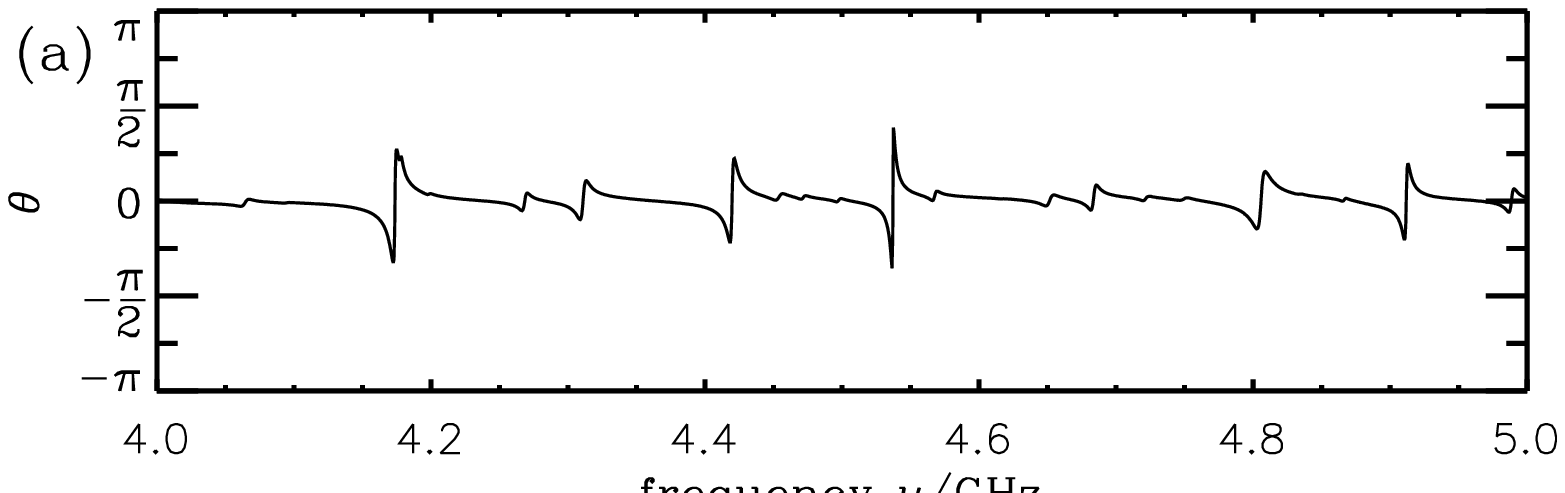}\hspace*{0.25cm} \includegraphics[height=1.9cm]{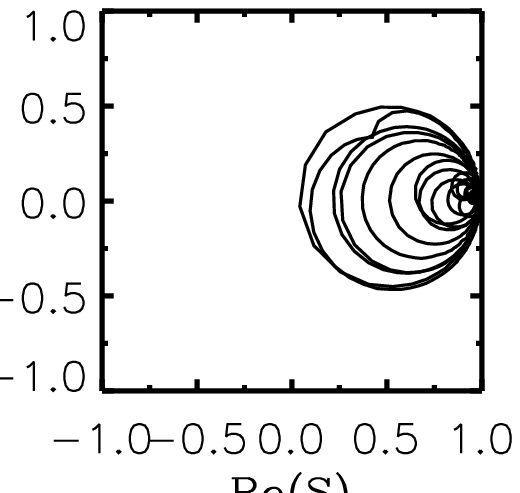}\\[1.5ex]
\includegraphics[height=1.9cm]{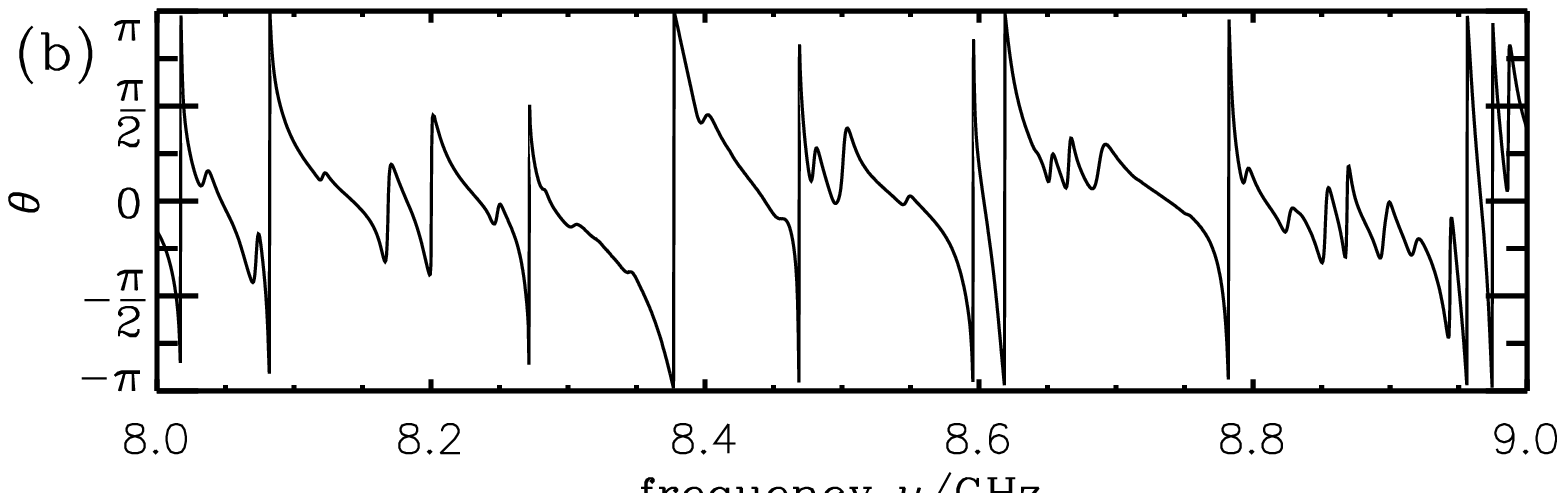}\hspace*{0.25cm} \includegraphics[height=1.9cm]{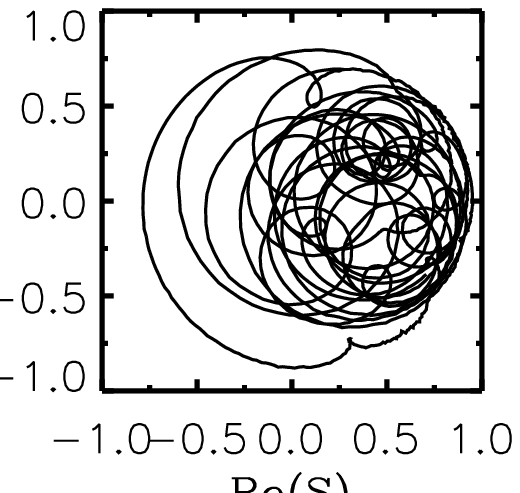}\\[1.5ex]
\includegraphics[height=1.9cm]{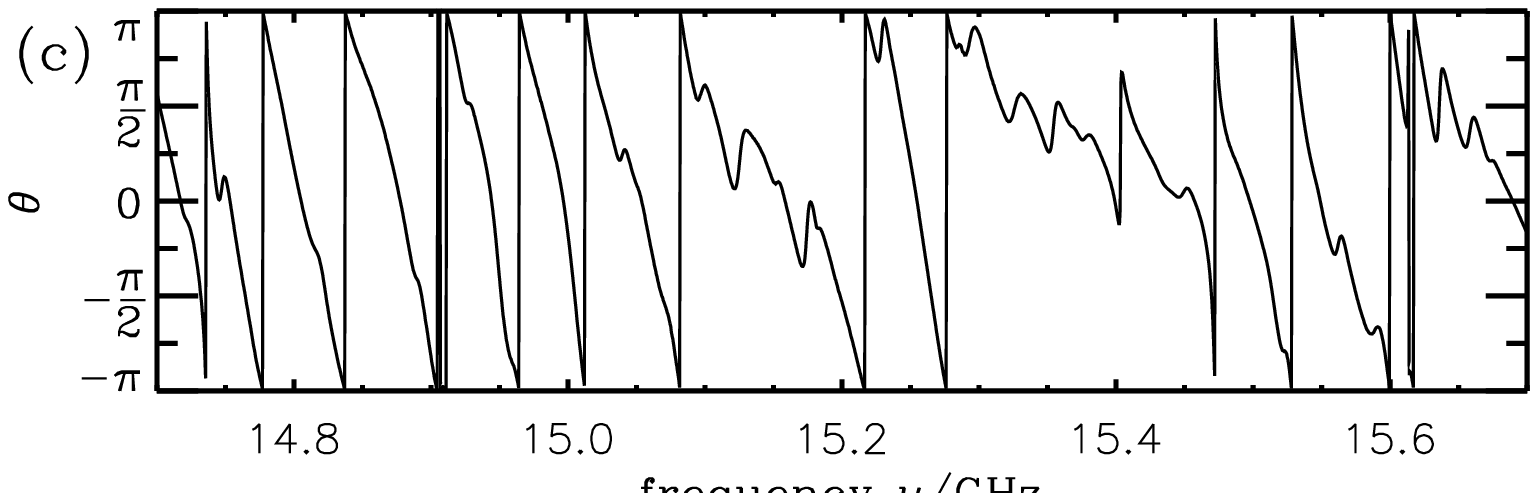}\hspace*{0.25cm} \includegraphics[height=1.9cm]{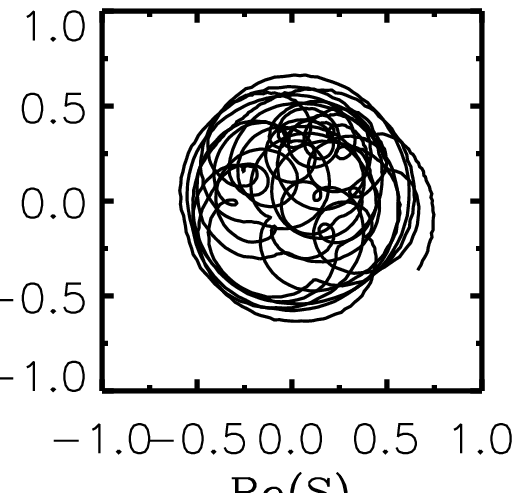}\\[1.5ex]
\includegraphics[height=1.9cm]{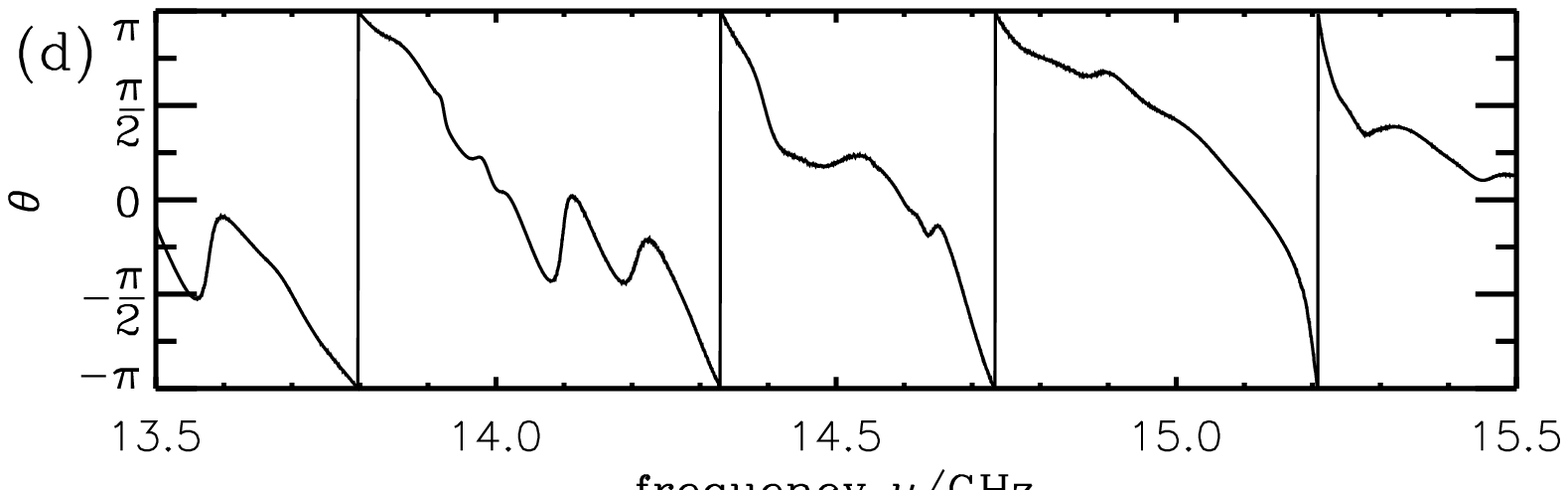}\hspace*{0.25cm} \includegraphics[height=1.9cm]{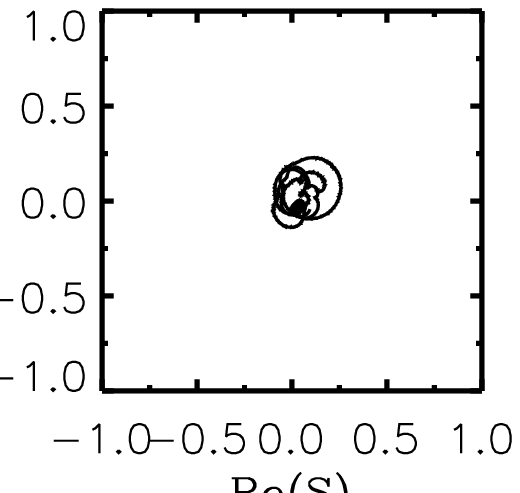}\\[1.5ex]
\caption{\label{fig:phase}
The phase $\theta$ of the $S$-matrix for different coupling and absorption regimes:
(a) $\gamma=0.56$,  $T_a=0.116$ (weak absorption-weak coupling),
(b) $\gamma=2.42$,  $T_a=0.754$,
(c) $\gamma=8.40$, $T_a=0.989$, and
(d) $\gamma=48.00$, $T_a=0.998$ (strong absorption-nearly perfect coupling).
The corresponding Argand diagrams are shown on the right hand side.}
\end{figure}

\begin{figure*}
\includegraphics[width=2\columnwidth]{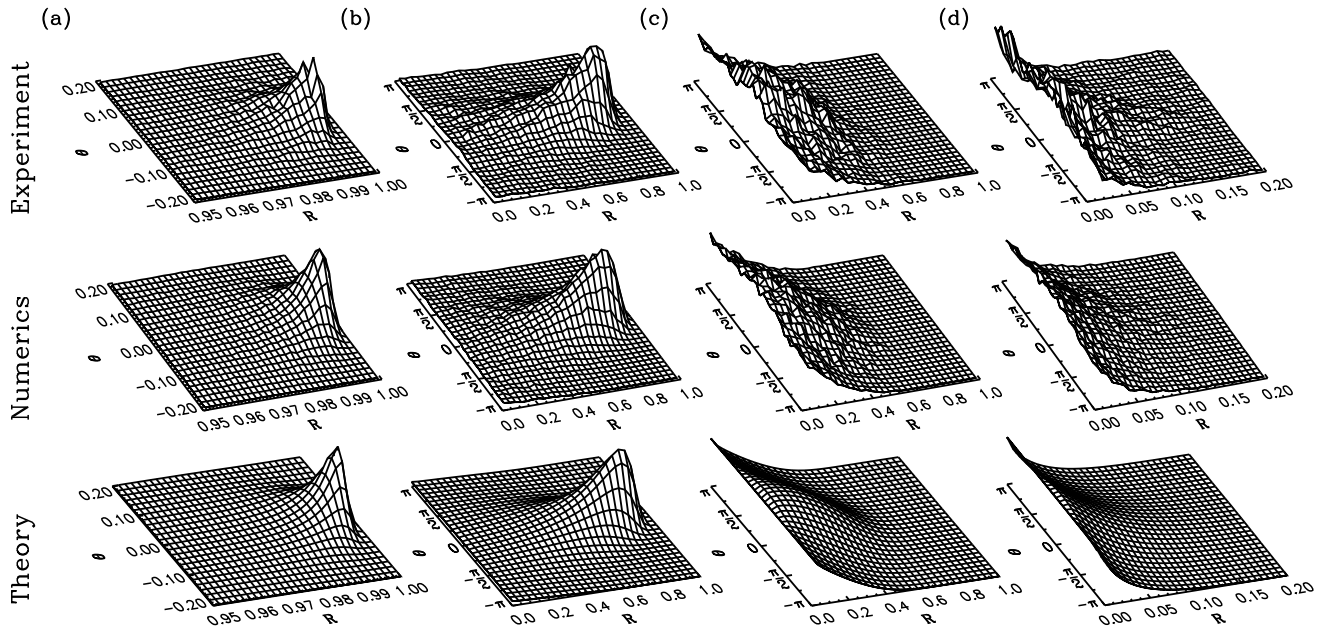}\\
\caption{Distribution of the $S$-matrix for the
same ranges of coupling and absorption of Fig.~\ref{fig:phase}.
The upper, central and lower rows correspond to experiment, numerics and
theory, respectively. Note the change of scales for $R$ and $\theta$.
}
\label{fig:PRT}
\end{figure*}

To derive the distribution $P_S(S)$ of the $S$-matrix in the presence of direct processes
and absorption for the one-channel case we consider the systems of
Fig.~\ref{fig:sketch}. In Fig.~\ref{fig:sketch}(a) a system with absorption strength
$\gamma$ and perfect coupling ($T_a$=1) is shown. Let us denote by
$S_0=\sqrt{R_0}e^{i\theta_0}$ the $S$-matrix describing the scattering in that
case. We assume that the distribution $P_{S,0}(S_0)$ of $S_0$ is known. Actually
it is sufficient to know the distribution of the reflection coefficient $P_{R,0}(R_0)$
under the assumption that $R_0$ and $\theta_0$ are still uncorrelated
when absorption is present.
As we mention above, the phase is uniformly distributed in the case
of perfect coupling and no absorption, i.e.,
\begin{equation}
P_{S,0}(S_0)=\frac{1}{2\pi}P_{R,0}(R_0).
\label{eq:P0}
\end{equation}
Just as for systems without absorption, the $S$-matrix of the system with direct
processes and absorption given by Eq.~(\ref{eq:para}) can be written in terms of the
$S$-matrix of the system with absorption but without direct processes, $S_0$, and the
$S$-matrix of the barrier $S_a$ that describes the non-ideal coupling of the antenna
(see Fig.~\ref{fig:sketch}(b)). As a model, we take the $S$-matrix of the barrier to be
\begin{equation}
S_{a}=\left(
\begin{array}{cc}
-\sqrt{1-T_a} & \sqrt{T_a} \\
\sqrt{T_a}  & \sqrt{1-T_a}
\end{array}
\right),
\label{Sbarrier}
\end{equation}
where $T_a$ is the transmission coefficient of the coupling barrier. The combination
rule of $S$-matrices gives the following relation between $S$ and $S_0$
\begin{equation}
S_0(S) = \frac{S-\langle S\rangle }
{1-\langle S\rangle S},
\label{transformation}
\end{equation}
where $\langle S\rangle= \sqrt{1-T_a}$. Then the distribution $P_S(S)$ for the system
including direct processes is $P_{S,0}(S_0)$ times the Jacobian of the transformation
(\ref{transformation}). The result is
\begin{eqnarray}
P_S(S) &=& \left|\frac{\partial (R_0, \theta_0) }{\partial(R,\theta)}\right| P_{S,0}(S_0)
           \nonumber \\
       &=& \left( \frac{1-\langle S \rangle^2}{|1-S \langle S \rangle|^2} \right)^2
           \frac{1}{2\pi} P_{R,0}(R_0),
\label{eq:P(R,theta)}
\end{eqnarray}
where $R_0=|S_0|^2$. The distribution $P_{R,0}(R_0)$ is known for
several cases \cite{Mel00,Bee01}:
\begin{equation}
P_{R,0}(R_0)= \left\{
\begin{array}{lll}
\frac{\exp[-\alpha/(1-R_0)]}{(1-R_0)^{3}}
\left[ A + B (1 - R_0 ) \right]; && \beta=2,  \\
C e^{-\alpha/(1-R_0)}/(1-R_0)^{2+\beta/2}  ;  &&\gamma \ll 1 \\
\alpha e^{- \alpha R_0 }; && \gamma \gg 1,
\end{array}
\right.
\label{P(R_0)}
\end{equation}
where $\alpha={\gamma \beta}/{2}$, $A={\alpha} \left(e^{\alpha}-1\right)$,
$B=( 1 +\alpha - e^{\alpha} )$, and $C=\alpha^{1+\beta/2}/\Gamma(1+\beta/2)$.
In the first case of Eq.~(\ref{P(R_0)}) the absorption can take any value,
whereas in the other cases $\beta$ can take any value.
As our experimental results are for a time reversal system we need the distribution
for $\beta=1$. An approximate distribution for $\beta=1$ was given in
Ref. \cite{Sav03}, that is valid, however, for intermediate and strong absorption only.
The following interpolation formula
\begin{equation}
P_{R,0}(R_0) =  C_\beta \frac{ e^{-\frac{\alpha}{1-R}}
}{(1-R)^{2+\beta/2}} \left[ A \alpha^{\beta/2-1} + B
(1-R)^{\beta/2} \right] \label{Suggestion}
\end{equation}
with $C_{\beta} =(A \Gamma(1+\beta/2, \alpha)/\alpha^2 + B e^{-\alpha}/\alpha)^{-1}$,
where $\Gamma(x,\alpha) =\int_\alpha^\infty t^{x-1}e^{-t}dt$ is the upper incomplete
Gamma-function. It satisfies all cases of Eq.~(\ref{P(R_0)}).
Notice that the theory presented here is rigorous for the imperfect
coupling or direct processes, only the ansatz given by Eq.~(\ref{Suggestion}) for the
absorption is heuristic.

The experimental setup was described in
Ref.~\cite{Men03}. The following two systems
were investigated: i) a half Sinai billiard and ii) a half Sinai billiard with a
microwave absorber attached to one side. To improve statistics, in both cases the
semicircle was moved along the wall in steps of 5\,mm to obtain more than 50
measurements. The complex $S$-matrix was measured with a vector network analyzer. We
investigated four different regimes also considered in
Ref.~\cite{Men03} ranging from weak absorption and
weak coupling ($\gamma=0.56$, $T_a=0.116$) to strong absorption and nearly perfect
coupling ($\gamma=48$, $T_a=0.998$). The one-channel case is realized due to the fact
that only a single antenna is attached with a radius much smaller than the wavelength.

In Fig.~\ref{fig:phase} we plot the phase $\theta$ of $S$ as a function of the
frequency and the corresponding Argand diagrams of the scattering matrix $S$ for
different coupling and absorption regimes. The experimental value of $T_a$ is
obtained directly from the mean value of the $S$-matrix by Eq.~(\ref{eq:coupling}).
The measured value of $\langle R\rangle$ then fixes the absorption strength $\gamma$
which decreases monotonically as a function of $\langle R\rangle$ and viceversa.

To test the theoretical and experimental results
we made numerical simulations based on random-matrix
theory using the Heidelberg approach. To obtain the numerical distributions we used
an ensemble of $S$-matrices constructed according to
\begin{equation}
S(E) = 1 - 2\pi i W^\dagger (E - H + i \pi W W^\dagger)^{-1} W,
\end{equation}
where $H$ is taken from the Gaussian orthogonal ensemble and $W$ gives the coupling
between the resonant modes of the cavity and the channel states, and $E$ is related to the electromagnetic frequency
$\nu$ by Weyl's formula. More details of the numerical method can be found in
Ref. \cite{Men03}.

\begin{figure}
\includegraphics[width=\columnwidth]{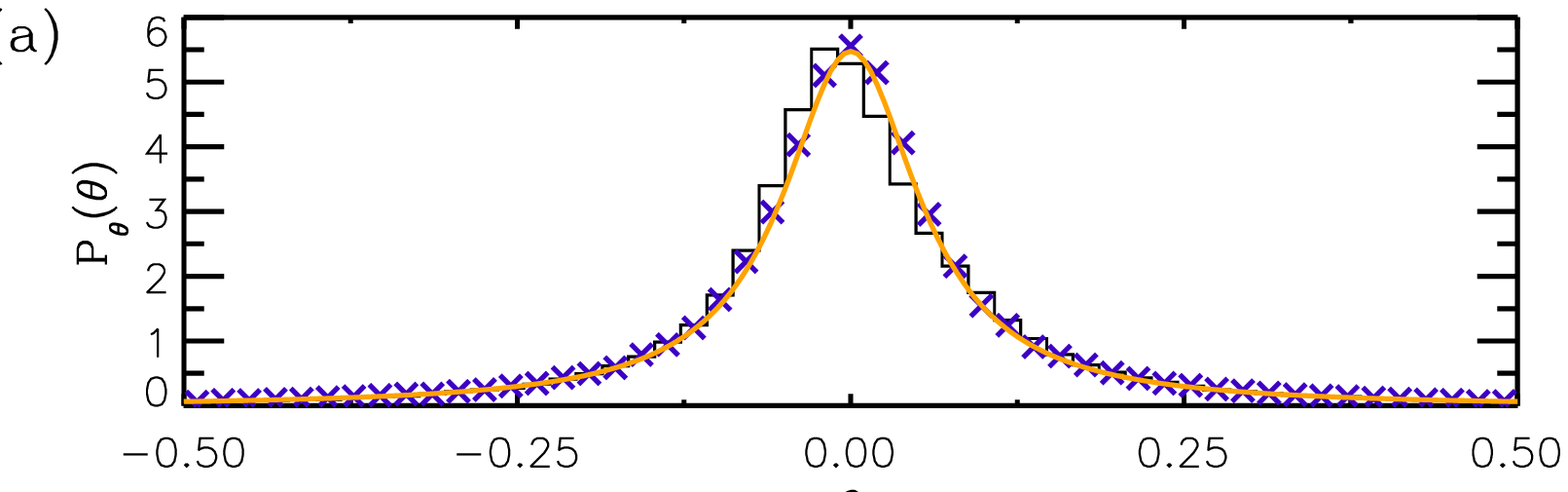}\\[0.25cm]
\includegraphics[width=\columnwidth]{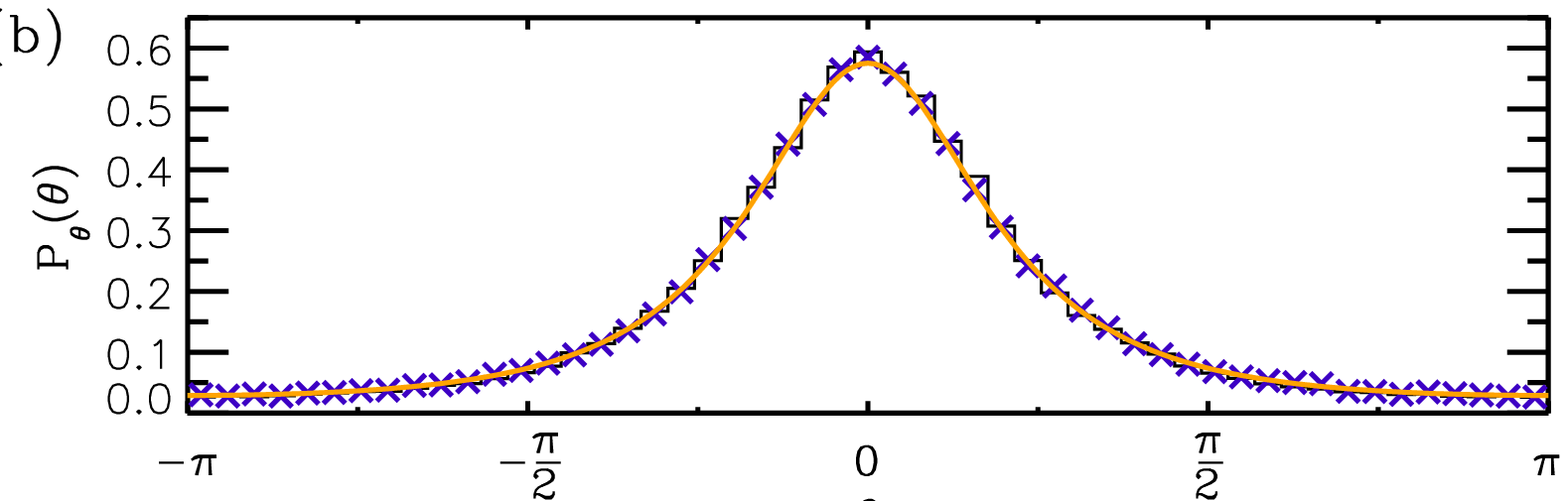}\\[0.25cm]
\includegraphics[width=\columnwidth]{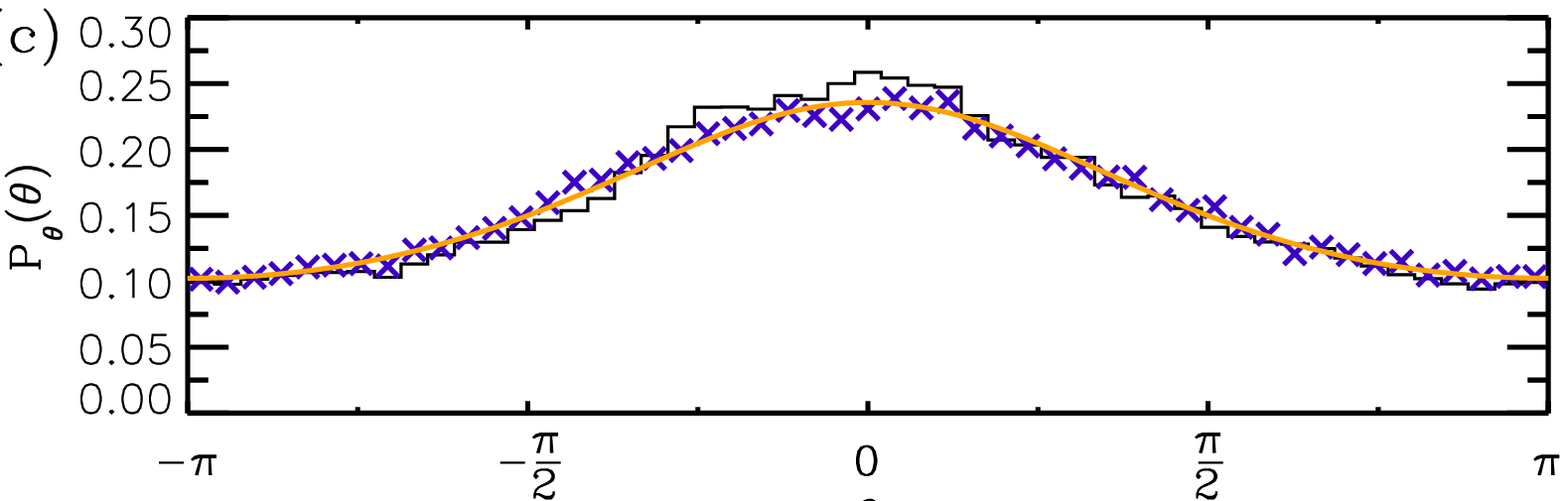}\\[0.25cm]
\includegraphics[width=\columnwidth]{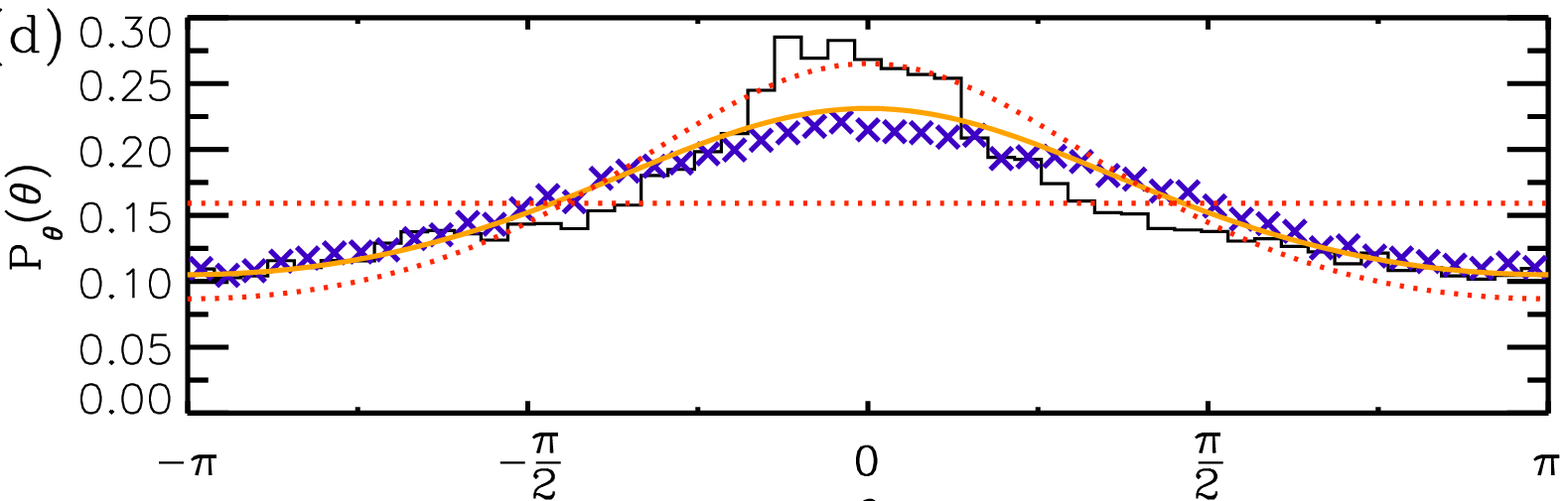}
\caption{\label{fig:PTheta}
Experimental distribution $P_{\theta,T}(\theta)$
(histogram) for the same regimes of coupling and
absorption as given in Fig.~\ref{fig:phase}.
Additionally the numerical (crosses) and the theoretical (solid line)
is plotted as well. In (d) the dotted curves correspond
to the coupling strengths $T_a$=0.996 and 1, respectively.}
\end{figure}

In Fig.~\ref{fig:PRT} we plot the experimental distributions of the $S$-matrix for
the same cases as in Fig.~\ref{fig:phase}. The numerical distributions are also
included in order to test the theoretical ones based on Eqs.~(\ref{eq:P0}),
(\ref{eq:P(R,theta)}), and (\ref{Suggestion}). An excellent agreement between
theory, numerics and experiment is found.
Finally, in Fig.~\ref{fig:PTheta} we show the distribution $P_\theta(\theta)$ for the
different regimes of absorption and antenna coupling. The numerical simulation is
included as well as the theoretical results obtained by integrating
Eq.~(\ref{eq:P(R,theta)}) numerically. Again an excellent agreement is found for
Fig.~\ref{fig:PTheta} (a)-(c). The deviations in the case of strong absorption and
nearly perfect coupling, $T_a$=0.998, (Fig.~\ref{fig:PTheta}(d)) are due to the fact
that the distribution $P_\theta(\theta)$ depends very sensitive on the coupling parameter
$T_a$. Within the investigated frequency range the coupling $T_a$ changes, which is
not taken into account in the theory. We additionally plotted the theoretical
distributions for $T_a$ =0.996 and 1.

In conclusion, we have shown experimental evidence of direct processes in the
scattering of chaotic microwave cavities in presence of absorption. This was done by
i) obtaining the experimental distribution of the $S$-matrix for various regimes of
absorption and antenna coupling in the one-channel case, ii) obtaining the same
distributions numerically using the Heidelberg approach, and iii) presenting a theory
that generalizes Poisson's kernel to include absorption.
Excellent agreement between theory, numerics and experiment was
obtained in all regimes of absorption and antenna coupling.
$P_{R,0}(R_0)$ as given by Eq.~\ref{Suggestion} is not rigorously derived for $\beta=1$.
As soon as a rigorously derived result becomes available
it can be plugged into Eq.~(\ref{eq:P(R,theta)}) to get the exact
distribution of the $S$ matrix for any absorption and antenna coupling.

This work was supported by DGAPA-UNAM, Conacyt-Mexico, and the DFG (Germany). MMM
thanks Centro Internacional de Ciencias A.~C. and Centro de Ciencias F\'{\i}sicas for
kind hospitality.
We thank C.~Lewenkopf, P. Mello and J.~Flores for useful comments.
We are especially indebted to D. Savin and Y. Fyodorov for suggesting this
improved version of Eq.~\ref{Suggestion} using time delay results from
Refs.~\cite{Sav01,Fyo03}.


\begin{thebibliography}{10}

\bibitem{Bee97}
C.\ W.\ J.\ Beenakker, Rev. Mod. Phys. {\bf 69},  731  (1997).

\bibitem{Stoe99}
H.-J.\ St\"ockmann, {\it Quantum Chaos}, (Cambridge University Press, Cambridge, 1999).

\bibitem{Mel04}P.\ A.\ Mello and N.\ Kumar, {\it Quantum Transport in
Mesoscopic Systems}, (Oxford University Press, Oxford 2004).

\bibitem{Men03}R.\ A.\ M\'endez-S\'anchez,
U.\ Kuhl, M.\ Barth, C.\ H.\ Lewenkopf and H.-J.\ St\"ockmann,
Phys. Rev. Lett. {\bf 91}, 174102 (2003).

\bibitem{Lop81}G.\ L\'opez, P.\ A.\ Mello and T.\ H.\ Seligman,
Z. Phys. A {\bf 302}, 351 (1981).

\bibitem{Sel83}T.\ H.\ Seligman in Proceedings Cali ELAF 1982 {\sl
Stochastic Processes Applied to Physics and other Related Topics }, eds.
A. Rueda and S. Moore World Scientific, Singapore 1983.

\bibitem{Mel85}P.\ A.\ Mello, P.\ Pereyra and T.\ H.\ Seligman,
Ann. Phys. {\bf 161}, 254 (1985).

\bibitem{Fri85}W.\ A.\ Friedman and P.\ A.\ Mello, Ann. Phys. {\bf 161},
276 (1985).

\bibitem{Dor92}E.\ Doron and U.\ Smilanzky, Nucl. Phys. A {\bf 545},
455 (1992).

\bibitem{Bro95}P.\ W.\ Brouwer, Phys. Rev. B {\bf 51}, 16878 (1995).

\bibitem{Sav01}D.\ V.\ Savin and  Y.\ V.\ Fyodorov and H.-J.\ Sommers,
Phys. Rev E {\bf 63}, 035202 (2001).

\bibitem{Acg03}G.\ B.\ Akguc and L.\ E.\ Reichl, Phys. Rev. E {\bf 67},
46202 (2003).

\bibitem{Mel00}E.\ Kogan, P.\ A.\ Mello, and He\ Liqun, Phys. Rev. E
{\bf 61},  R17  (2000).

\bibitem{Bee01}C.\ W.\ J.\ Beenakker and P.\ W.\ Brouwer, Physica E
{\bf 9},  463  (2001).

\bibitem{Sav03}D.\ V.\ Savin and  H.-J.\ Sommers,
Phys. Rev E {\bf 68}, 36211 (2003).

\bibitem{Sav04}D.\ V.\ Savin and  H.-J.\ Sommers,
Phys. Rev E {\bf 69}, 35201 (2004).

\bibitem{Fyo03}Y.\ V.\ Fyodorov, JETP Lett. {\bf 78}, 250 (2003).

\bibitem{Fyo04}Y.\ V.\ Fyodorov and A.\ Ossipov, Phys. Rev. Lett.
{\bf 92}, 084103 (2004).

\bibitem{Dor90}E.\ Doron, U.\ Smilansky and A.\ Frenkel,
Phys. Rev. Lett. {\bf 65}, 3072 (1990).

\bibitem{Lew92}C.\ H.\ Lewenkopf, A. M\"uller and E. Doron,
Phys. Rev. A {\bf 45}, 2635 (1992).

\bibitem{Sch03}R.\ Sch\"afer, T.\ Gorin, T.\ H.\ Seligman,
and H.-J.\ St\"ockmann, J. Phys. A {\bf 36}, 3289 (2003).

\bibitem{Sch01}H.\ Schanze, E.\ R.\ P.\ Alves,
C.\ H.\ Lewenkopf, and H.-J.\ St\"ockmann,
Phys. Rev. E {\bf 64}, 065201(R) (2001).

\bibitem{Hem}S.\ Hemmady, X.\ Zheng, E.\ Ott,
T.\ M.\ Antonsen and S.\ M.\ Anlage, cond-mat/0403225.

\end{thebibliography}
\end{document}